\documentclass[3p,times,twocolumn]{elsarticle}

\usepackage{ecrc}

\volume{00}

\firstpage{1}

\journalname{Nuclear Physics B Proceedings Supplement}

\runauth{A.~Rubbia}

\jid{nuphbp}

\jnltitlelogo{Nuclear Physics B Proceedings Supplement}

\usepackage{amssymb}

\usepackage[figuresright]{rotating}

\begin{document}

\begin{frontmatter}



\dochead{}

\title{Future liquid Argon detectors}

\author[ETHZ]{A.~Rubbia}
\address[ETHZ]{ETH Zurich, 101 R\"amistrasse, CH-8092 Zurich, Switzerland}

\begin{abstract}
The Liquid Argon Time Projection Chamber offers an innovative technology
for a new class of massive detectors for rare-event detection. 
It is a precise tracking device that allows three-dimensional
spatial reconstruction with mm-scale precision of the morphology of
ionizing tracks with the imaging quality of a ``bubble chamber'', 
provides $dE/dx$ information with high sampling rate, and 
acts as high-resolution calorimeter for contained events.
First proposed in 1977 and after a long maturing process, its holds today the 
potentialities of opening new physics opportunities by
providing excellent tracking and calorimetry performance at  the relevant multi-kton
mass scales, outperforming other techniques. 
 In this paper,
we review future liquid argon detectors presently
being discussed by the neutrino physics community.
\end{abstract}

\end{frontmatter}


\section{Introduction}
\label{sec:introduction}

The Liquid Argon Time Projection Chamber (LAr TPC)~\cite{Rubbia:1977zz} 
is the  successful marriage between the ``gaseous'' TPC chamber and 
``the liquid argon calorimeter'', to obtain a combined
dense and very fine-grained three-dimensional 
tracking device with local $dE/dx$ information and a homogenous full sampling calorimeter.

On one hand, the ``gaseous'' TPC~\cite{marx:46} has been a seminal development in particle tracking detectors.
Unlike any other tracking detector, it provided superb capabilities for tracking of charged particles
coupled with electric charge (when embedded 
in a B-field), momentum and type identification through energy-loss measurements.
It was the only tracking detector where for each point on the track, x-, y- and z-coordinates were 
measured simultaneously and with very high spatial resolution. This played
a particularly important role in pattern recognition of complex reactions.

On the other hand, the invention of
the homogeneous liquid argon ionization chamber 
has allowed
the energy measurement by total absorption with an 
unprecedented precision. 
As was pointed out in the pioneering work~\cite{Willis:1974gi}, liquid argon is a medium that satisfies 
calorimetry requirements better than many other materials.
All the energy is ideally 
converted into ionisation through the development of
the cascade showers (electromagnetic and hadronic). The charge
is drifted across a gap 
and picked up electronically with sensitive charge preamplifiers. 
The integrated collected charge
yields, once calibrated, the incoming particle energy.
Although argon is a non-compensating medium,
the imaging properties of the LAr TPC would in principle
allow to pass events through sophisticated energy-flow algorithms
and perform ``offline'' $e/\pi$ compensation for
best performance achievements.

As a result, the combined and excellent tracking-calorimeter performances of the LAr TPC
and a design that can be extrapolated to giant multi-kton scale~\cite{Rubbia:2004tz,Rubbia:2009md},
allow to contemplate next-generation massive neutrino detectors with higher signal efficiency and 
effective background discrimination compared to other techniques, with optimal
conditions to detect, identify and precisely measure a very large class of 
neutrino interactions from the lowest energies in the MeV range to complex
multi-GeV events.

Compared to the current state-of-the art represented by SuperKamiokande,
multi-kton liquid argon detectors will open new so-far-unexplored physics domains thanks
to the improved performance, and they hence represent a complementary
approach to gigantic Water-Cerenkov detectors like HyperKamiokande~\cite{Nakamura:2000tp,Abe:2011ts}.

\section{Physical parameters and challenges}
Some relevant physical properties of argon are summarised in Table~\ref{tab:argpar}.
\begin{table}
\centering
\begin{tabular}{|c|c|c|c|c|c|c|}
\hline
Atomic number & 18 \\
Molecular weight & 39.948 g$/$mol \\
Natural concentration & 0.934\% of air \\
Melting point & 83.4K at 1 atm \\
Boiling point & 87.3K at 1 atm \\
Triple point & 83.8K at 0.687 bar \\
Liquid density (at 83.7K) & 1392.8 kg$/$m$^3$\\
Latent heat (1 atm) & 160.81 kJ$/$kg\\
Dielectric constant & 1.5 \\
Electric breakdown & 1.1--1.4~MV$/$cm \\
\hline
$dE/dx$ for m.i.p. & 2.12~MeV$/$cm \\
Ionisation energy $W_e$ ($E=\infty$) & 23.6 eV \\
Excitation energy $W_\gamma$ ($E=0$) & 19.5 eV \\
Radiation length $X_0$ & 14~cm \\
$\gamma$ pair production length $(9/7)X_0$ & 18~cm \\
Moli\`ere radius & 9.28~cm \\
Nuclear interaction length $\lambda_{int}$& 84~cm \\
Critical energy & 30 MeV \\
\hline
\end{tabular}
\caption{Some physical parameters of argon.}
\label{tab:argpar}
\end{table}
Liquid Argon (87~K, 1 bar) is a high density, rather cheap and
easily accessible by-product
of the air liquefaction process. 
The concept of the LAr TPC is illustrated in Figure~\ref{fig:larttpcconcept}.

Quasi-free electrons from
ionising tracks are easily drifted in LAr with a
drift velocity $v_d\sim1.6-2$~mm/$\mu$s
with a modest electric field
of $0.5-1~$kV$/$cm. In comparison,
the drift velocity of ions is totally negligible. The electron cloud diffusion
is rather small, as the spatial spread of the cloud given by $\sigma\simeq \sqrt{2Dx/v_d}$ (with
$D=D_L, D_T$ are the longitudinal
and transversal diffusion coefficients $D_L\approx$4cm$^2$/s, $D_T\approx$13cm$^2$/s), 
is
at the level of $\approx$mm after several meters of drift path $x$. 
For an ultimate drift of 20~m at 1~kV/cm,  the longitudinal(resp. transversal) diffusion
is 3~mm(resp. 5~mm). Beyond this drift distance, the images will be
significantly distorted compared to the canonical imaging pitch. In order to create
the image of events in the chamber volume, the ionisation electrons
are drifted towards readout planes segmented
in individual electrodes with a sampling pitch of 3~mm or $\approx 2\%X_0$.
The drift motion induces currents that are continuously
shaped and digitised by the readout electronics. Signal
waveforms are processed offline by reconstruction and
pattern recognition algorithms to determine the features
of the events.

The LAr scintillation
properties are also excellent with about 40000 photons/MeV at 0.5~kV/cm, 
although the light is primarily localised around 128~nm (DUV) and hence cannot
be directly efficiently detected by photomultipliers (PMT) as it is does not traverse
glass or even crystal windows. In order to be directly sensitive, PMTs
with a $MgF_2$ window can be employed. A practical solution with glass window
PMTs
consists in coating surfaces of the detector and of the PMTs with wavelength
shifters, such as e.g. TPB~\cite{Boccone:2009kk}. The scintillation light can
provide the absolute time of an event ($T_0$), a prompt trigger, and also pulse shape
discrimination.

Several technical challenges have limited the use and widespread
of the LAr TPC technology.
The recent progresses and achievements are however
transforming pioneering R\&D efforts into realistic designs for very large detectors.
We list some issues relevant to the implementation of the LAr TPC
at the largest possible scale:

\begin{figure}
\begin{center}
	\resizebox{\linewidth}{!}{\includegraphics{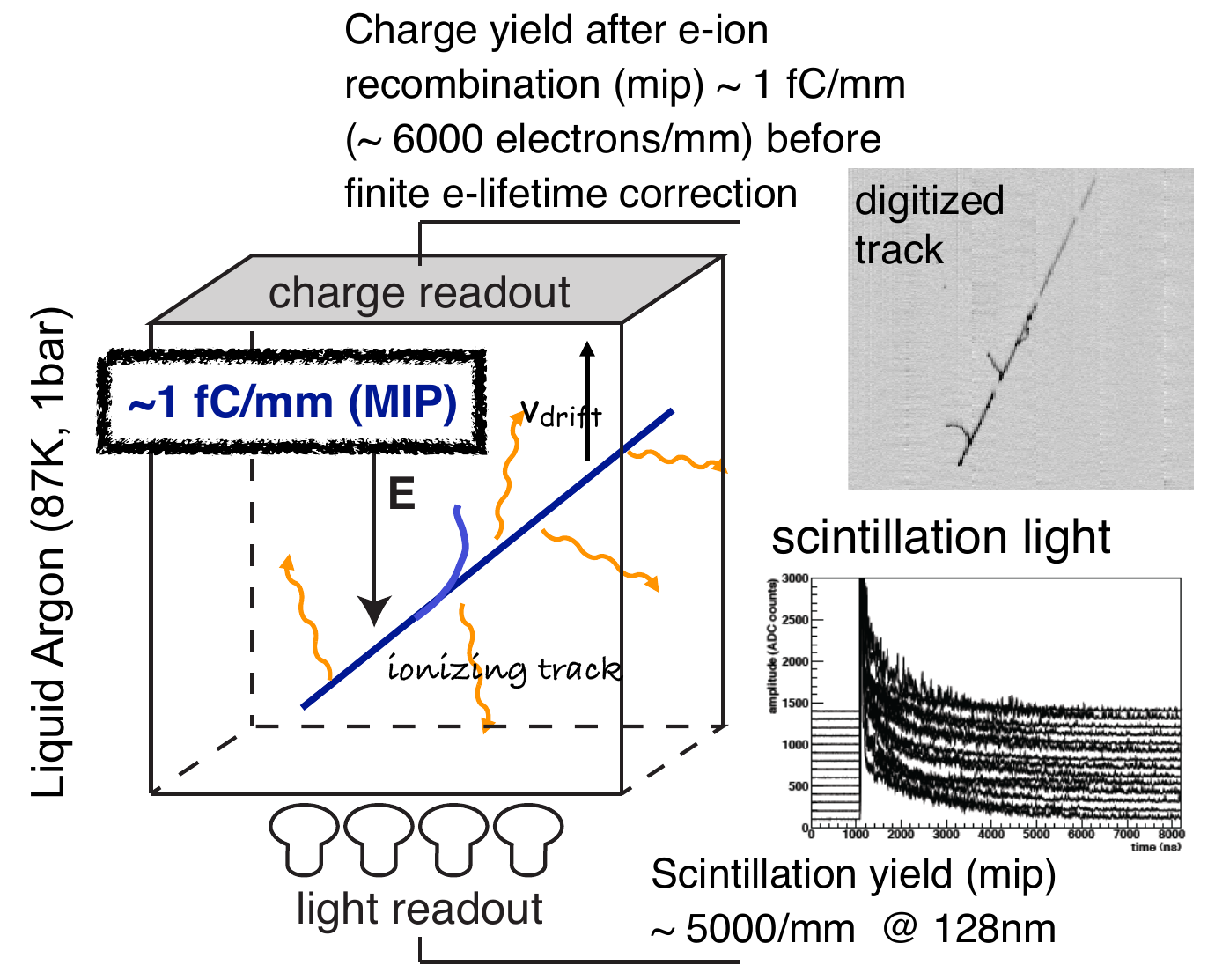}}
	\caption{Illustration of the LAr TPC detector concept.}
	\label{fig:larttpcconcept}
\end{center}
\end{figure}

\begin{figure}
\begin{center}
	\resizebox{0.8\linewidth}{!}{\includegraphics{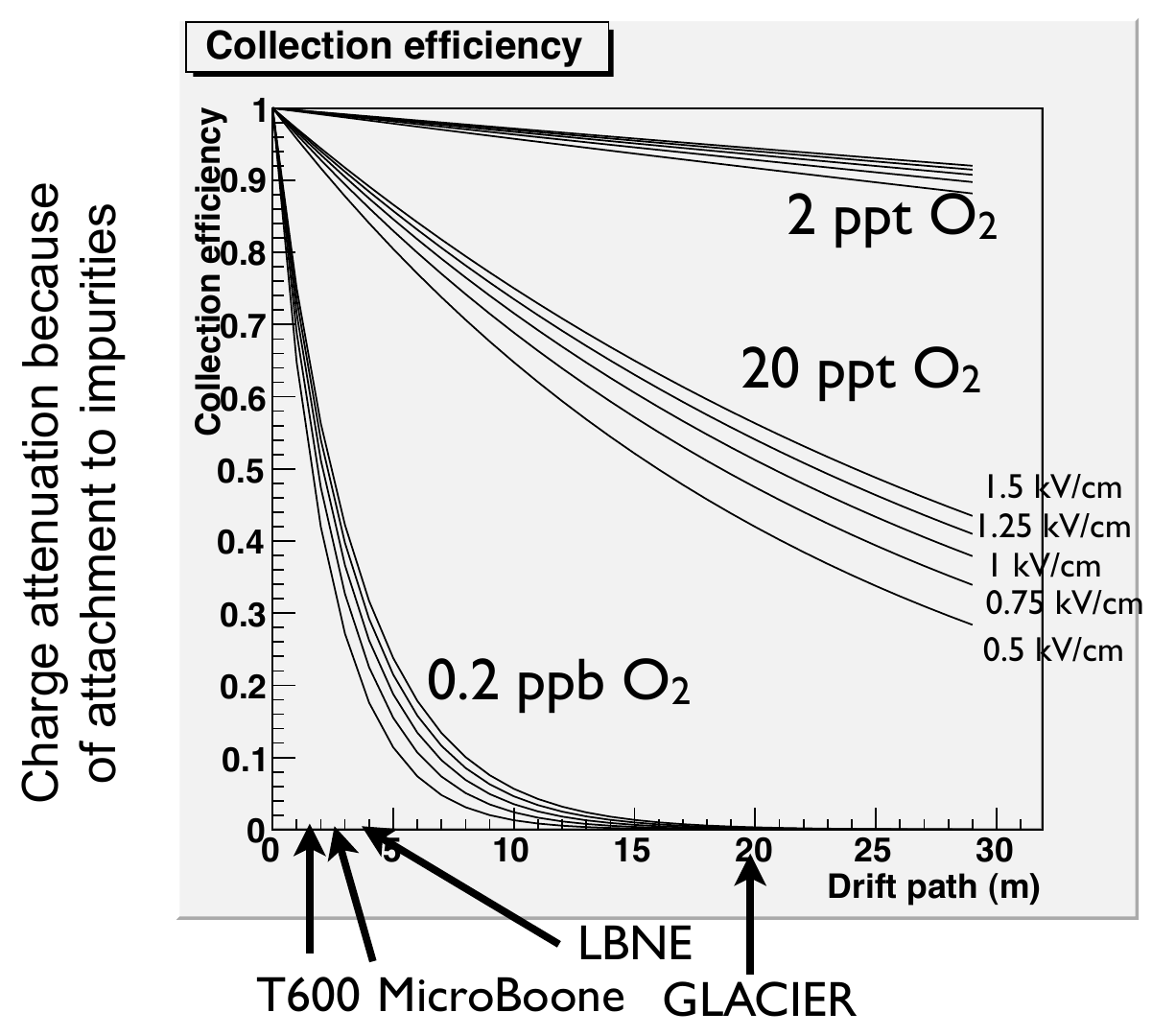}}
	\caption{Expected ionisation charge attenuation due to attachment
	to impurities as a function of the drift path for 0.2~ppb, 20~ppt and 
	2~ppt Oxygen-equivalent
	impurity levels and electric fields in the range 0.5--1.5~kV/cm. 
	The arrows indicate the drift length of
	the ICARUS T600, MicroBOONE, LBNE and LAGUNA-LBNO GLACIER.}
	\label{fig:larimpurity}
\end{center}
\end{figure}

\begin{enumerate}

\item The ionisation charge yield after e-ion
recombination for a m.i.p. particle is about 1~fC or 6000~$e^{-}$
per mm of track.
The lack of charge amplification in the liquid requires
fC-level charge sensitive preamplifiers and necessitate
to minimise the  capacitance load from the detector $C_{detector}$
and cables $C_{cables}$. In order to maintain
a signal-to-noise (S/N) ratio for m.i.p. particles, the preamplifiers
must typically have a noise $<1500$~ENC for $(C_{detector}+C_{cables})<400$~pF.

\item The  long drift path requires an ultra-high level of purity in the medium.
The expected ionisation charge attenuation due to attachment
to impurities as a function of the drift path for various oxygen-equivalent
impurity levels and electric fields is shown in Figure~\ref{fig:larimpurity}.
The arrows indicate the drift length of
the ICARUS T600~\cite{Amerio:2004ze}, 
MicroBOONE~\cite{Chen:2007ae}, LBNE~\cite{Akiri:2011dv} 
and LAGUNA-LBNO GLACIER~\cite{Rubbia:2009md,Rubbia:2010zz}.
The  free of electro-negative molecules (like $O_2$, $H_2O$, etc.) must
reach a level below 100~ppt $O_2$ level. An impurity level 
of $<30$~ppt $O_2$-equivalent is needed to obtain an electron lifetime
greater than 10~ms.
Compared to commercially
available bulk liquid argon deliveries which typically contain ppm-level
purities, the goal is to reduce those impurities by a factor $10^4$--$10^5$
before filling the main vessel tank. 
Excellent purity has been reproducibly achieved in various setups always relying on commercially available techniques, of various sizes and capacities, and should not pose a 
show-stopper for long drift paths.

\item Several independent groups performed numerical simulations and 
concluded that the vacuum evacuation of the main detector volume 
could be avoided for larger detectors,
thanks to 
(1) a more favourable surface / volume ratio for larger volume 
(also larger volumes are less sensitive to micro-leaks),
(2) a purification from ppm to $<<$ 1 ppb is anyhow needed
since the initial purity of argon when delivered is typ. ppm $O_2$ (see above),
and (3)  the outgassing of material is mostly from hot components
and impurities ÒfrozenÓ at low temperature.
GAr flushing and purging were shown to be effective ways to remove air and impurities.
Purging on 6~m$^3$ volume has been successfully demonstrated~\cite{Curioni:2010gd}.
The piston effect was seen in gas and the impurities
reached 3~ppm $O_2$ after several volumes exchange.
The Liquid Argon Purity Demonstrator (LAPD)
with a size of 30 ton has reach
a first milestone, obtaining a better than 3~ms electron lifetime in a 
large non-evacuated vessel~\cite{Rebel:2011zzc}.

\item In order to avoid contamination during
operation, large ultra high vacuum systems (leak rate $<10^{-9}$~mbar lt/s)
have been successfully implemented. Microleaks might
be the limiting factor to purity and hence long drift paths.

\item Long drift paths and high drift fields imply very high voltages on the cathode.
Voltages up to 150~kV have been reached. Immersed voltage multipliers
have been successfully developed and operated which could be
extrapolated to MV~\cite{Horikawa:2010bv,Badertscher:2012dq},
such as to allow 20~m drift distances. However, the long-term stability 
against
discharges of such high voltages remains to be demonstrated.

\item In the single phase configuration, wire chambers with typical wire
length $>9$~m at cryogenic temperature
pose  mechanical challenges. The breaking of a single wire can
be dramatic. 

\item The operation at $T=87$~K
of large cryogenic systems
create technical challenges that can be addressed 
by storage
and process techniques
developed by the petrochemical industry~\cite{Rubbia:2004tz}.

\item A novel method of readout operation called the
the double phase LAr LEM TPC with adjustable gain, has been successfully
developed and operated~\cite{Rubbia:2004tz}. See Section~\ref{sec:larlemtpc}.
This multiplication produces
a significant advantage over traditional systems. These devices are 
also more robust than the traditional wires with no chance 
of a channel breaking due to mechanical stress.
\end{enumerate}

\section{Double phase readout: the LAr LEM TPC}
\label{sec:larlemtpc}
The Liquid Argon Large Electron Multiplier Time Projection Chamber (LAr LEM-TPC) is 
a novel kind of double phase (liquid-vapor) noble gas TPC with adjustable gain~\cite{Rubbia:2004tz,Rubbia:2009md,Badertscher:2012dq,Badertscher:2008rf,Badertscher:2009av,Badertscher:2010fi,Badertscher:2010zg}. 
Thanks to the gain, each volumetric pixel (voxel) is reconstructed 
with better signal-to-noise ratio and consequently a lower energy deposition threshold is possible
than in single-phase.
In addition charge amplification 
reduces the impact of charge dilution due to the longitudinal diffusion of the electron cloud along the drift paths, 
and can be used to compensate for potential charge losses due to electronegative impurities diluted in the liquid argon. 
Therefore, this technology is very promising and possibly represents the ``enabling'' choice
for the realisation of the next generation underground
detectors for neutrino physics with imaging technique, in all cases with
a significant improvement of the imaging quality compared to 
the single phase liquid argon TPC~\cite{Badertscher:2009av}.

\begin{figure}
\begin{center}
	\resizebox{0.9\linewidth}{!}{\includegraphics{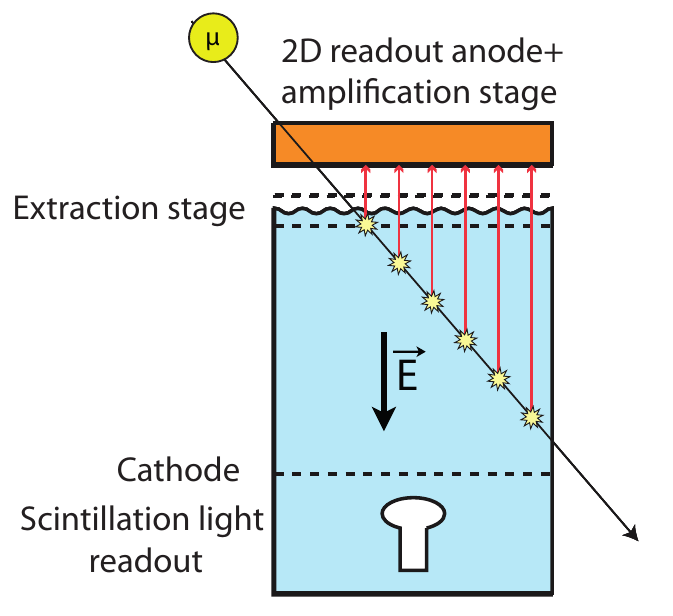}}
	\caption{Concept of the double phase LAr LEM-TPC with adjustable gain.
	After the ionisation electrons have been extracted from the liquid to 
	the gas phase, the charge is amplified via Townsend multiplication in high
	electric field present in macroscopic holes,
	before it is collected by the two-dimensional readout anode.}
	\label{fig:doublephaselartpc}
\end{center}
\end{figure}

In the LAr LEM TPC, the charge produced by ionizing particles in the liquid is drifted 
towards the liquid-vapor interface, where electrons are extracted 
to the vapor phase by means of an appropriate
electric field produced by two grids. See Figure~\ref{fig:doublephaselartpc}.
In the vapor phase, Townsend avalanche takes place in the high electric field regions confined 
in the LEM holes, similar to the situation of the Gas Electron Multiplier (See Ref. in~\cite{Badertscher:2010zg}). 
The LEM is a macroscopic hole electron multiplier built with standard PCB techniques. 
A HV of typically 3~kV is applied across the two faces creating a strong electric field of 30~kV/cm 
in the LEM holes leading to multiplication of electrons by avalanches. 
The amplified charge is collected by a set
of segmented electrodes, on which signals are induced.
In order to obtain a complete 3D-spatial reconstruction of ionizing events, moving charges have to induce signals on (at least) two complementary X-Y
sets of electrode strips. 
This  is achieved thanks to the use of a projective 2D 
charge X-Y-readout, providing two independent views with 3~mm effective pitch. 
Based on the GEM 2D readout concept (see Ref. in~\cite{Badertscher:2010zg}), it consists of two perpendicular sets of strips with a 50~$\mu$m thick Kapton spacer in-between. The drifting charge is collected on both sets of strips, based on the principle of balanced charge sharing. 
The Z~(drift) coordinate is derived by measuring the primary electrons drift time. 
%

\begin{figure}
\begin{center}
	\resizebox{\linewidth}{!}{\includegraphics{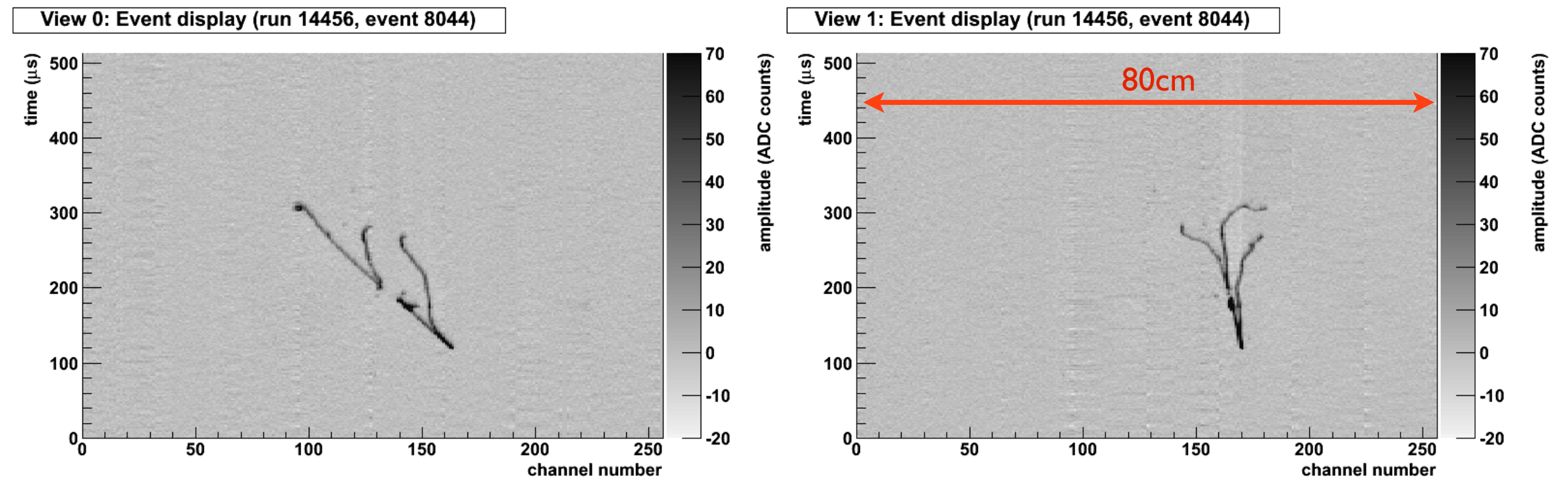}}
	\caption{Cosmic ray events collected with the 40$\times$80~cm$^2$
	double phase LAr LEM-TPC prototype.
	The two pictures correspond to two
perpendicular independent views of the event (labelled View 0 and 1). The vertical axes correspond to the drift time and
the horizontal axes to the channel number.}
	\label{fig:doublephasevents}
\end{center}
\end{figure}

A  high signal-to-noise ratio  can be reached in the LAr LEM-TPC thanks to the gas amplification stage.
This significantly  improves the 
event reconstruction quality 
with a lower energy deposition threshold and a better resolution per volumetric pixel (voxel) 
compared to a conventional single-phase LAr TPC 
\cite{Badertscher:2009av}.
Figure~\ref{fig:doublephasevents} shows beautiful
cosmic ray events collected with a 40$\times$80~cm$^2$
double phase LAr LEM-TPC prototype~\cite{Badertscher:2012dq}.
\begin{figure}
\begin{center}
	\resizebox{\linewidth}{!}{\includegraphics{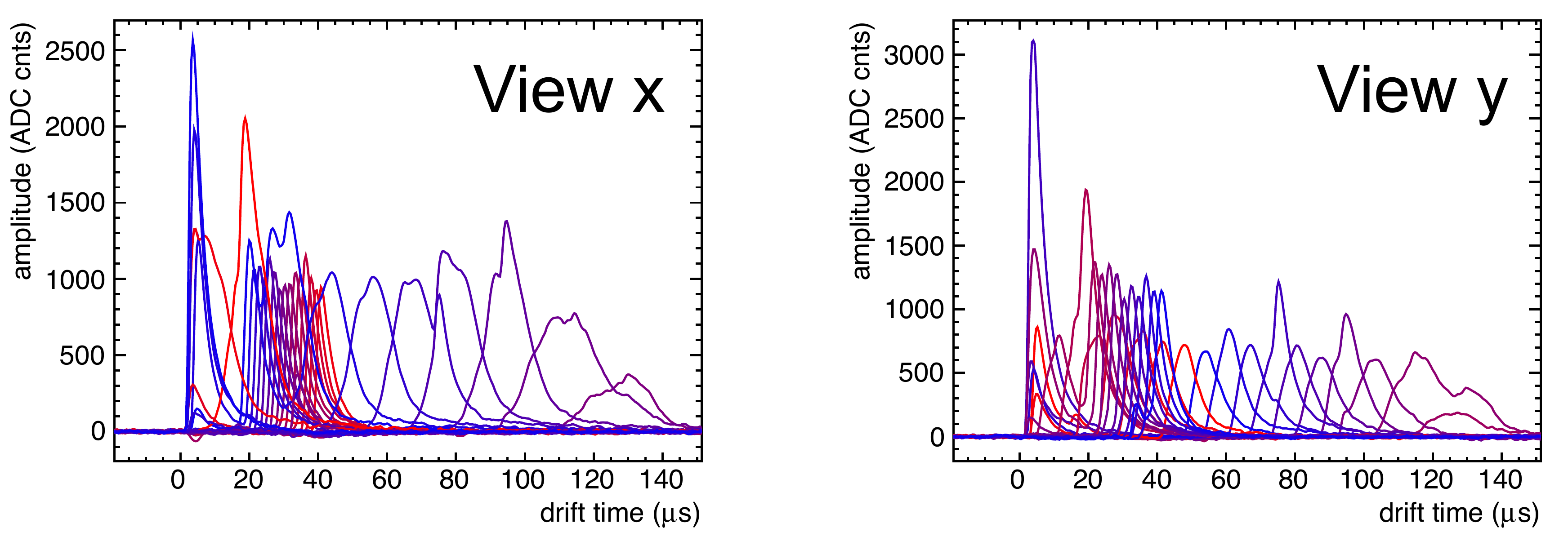}}
	\caption{Signal waveforms of cosmic ray events collected with the 40$\times$80~cm$^2$
	double phase LAr LEM-TPC prototype.}
	\label{fig:doublephasesignals}
\end{center}
\end{figure}
The two pictures correspond to two
perpendicular independent views of the event (labelled View 0 and 1). The vertical axes correspond to the drift time and
the horizontal axes to the channel number.
Signal waveforms from the same prototype are displayed in Figure~\ref{fig:doublephasesignals}.
The excellent S/N is impressive.
The charge amplification can
compensate for potential losses of signal-to-noise due to the charge diffusion and
attachment to electronegative impurities diluted in LAr, 
which both become more important as the drift length increases. The collection-only 
readout mode (avoiding the use of induction planes) is also an important asset
in the case of complicated topologies, like e.g. in electromagnetic or hadronic showers.

\section{Overview of detector performance}
The LAr TPC offers excellent combined tracking and calorimetric performance.
The detector and physics performance has been studied with small-scale
prototypes and detailed detector simulations.
The 3-dimensional tracking has a mm-scale spatial resolution with local
$dE/dx$ measurement at each point. It is fully sensitive medium, with a
$\approx 2\%X_0$ sampling rate for 3~mm pitch, and excellent energy resolution for
contained events. 

\begin{figure}
\begin{center}
	\resizebox{0.9\linewidth}{!}{\includegraphics{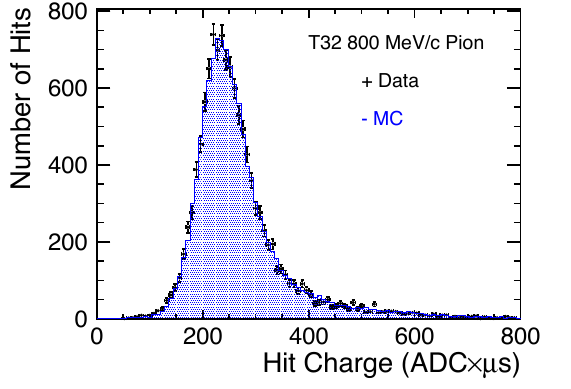}}
	\vspace{-0.3cm}
	\caption{$dE/dx$ response measured with the 250~L chamber exposed
	to the tagged low-momentum kaon test-beam 
at J-PARC.}
	\label{fig:t32dedx}
\end{center}
\end{figure}

The ability to reproduce well the $dE/dx$ performance has
been demonstrated with a 250~L chamber
exposed to a tagged low-momentum kaon test-beam 
at J-PARC~\cite{Araoka:2011pw}.
Figure~\ref{fig:t32dedx} shows the
$dE/dx$ response measured with the 250~L chamber exposed
	to the tagged low-momentum kaon test-beam 
at J-PARC. The data is very well described by the 
Birks law:
\begin{equation}
Q=A\frac{Q_0}{1+(\kappa/\epsilon)\times(dE/dx)\times(1/\rho)}
\end{equation}
where $A=0.8$, $\kappa=0.0486$~kV/cm g/cm$^2$/MeV,
$\epsilon$ the electric field, and $\rho$ the density.
As a consequence of the good  understanding of tracking
and in particular of the $dE/dx$ response, we anticipate that
particle identification combining range, $dE/dx$ and
imaging information should provide excellent $e/\pi^0$ separation
with typically $\epsilon \sim 90\%$ efficiency for a rejection
factor $>100$. 

The calorimetric response is less well known.
The reconstruction of Michel electrons form stopping muon decay sample
was studied in details and yields~\cite{Amoruso:2003sw}:
\begin{equation}
\frac{\sigma_e}{E}\simeq \frac{11\%}{\sqrt{E(MeV)}}\oplus 4\%
\end{equation}
The hadronic calorimetry has never been studied in a test-beam.
From Monte-Carlo simulations, the resolution for electromagnetic
showers is expected to be:
\begin{equation}
\frac{\sigma^{MC}_{em}}{E} \simeq \frac{3\%}{\sqrt{E}}\oplus 1\%
\end{equation}
where the constant term is dominated by the noise
and reconstruction effects, while for hadronic showers~\cite{Stahl:2012exa}
\begin{equation}
\frac{\sigma^{MC}_{had}}{E} \simeq  \frac{15\%}{\sqrt{E}}\oplus  B
\end{equation}
where $B \sim 5\%$ with GEANT3 and $B\sim10\%$ with GEANT4.
An extensive campaign on a test beam to study calorimetry 
would represent a great source of information to resolve these
discrepancies and improve the calorimetric understanding of
the LAr TPC, in particular in view of the precision measurements
to be performed in the future at long baseline for oscillation
measurements and CP-violation searches.

Similarly, a precise reconstruction of the incoming neutrino 
energy over a wide range of energies will be an important asset.
In a LAr TPC different methods can be used (see~\cite{Stahl:2012exa}).
Using the calorimetric approach, one can apply ``energy conservation'' and
sum the deposited energies of all outgoing particles,
obtained by summing the $dE/dx$ measurements along each ionising
track to obtain the associated kinetic energies $T\equiv \int (dE/dx) dx$.
One should identify final state particles in order to take into account their
rest masses. This method is less sensitive to Fermi motion than other
methods. Detailed full simulations of $\nu_e$~CC events
between 0 and 10~GeV show an excellent neutrino energy resolution 
$(E_{\nu}-E_{reco})/E_{\nu}$ of 8.4\% RMS~\cite{Stahl:2012exa}.

Kinematical reconstruction of quasi-elastic 
neutrino interactions has been published with 
ICARUS~50L data obtained at the CERN WANF~\cite{Arneodo:2006ug}.
An inclusive cross-section measurement has been
obtained with the ArgoNEUT data~\cite{Anderson:2011ce}.
Many phenomenological studies were published, for instance on
proton decay sensitivities~\cite{Bueno:2007um},
atmospheric neutrino detection (in particular tau appearance)~\cite{Bueno:2004ty,Campanelli:2000we},
supernovae core collapse neutrinos detection~\cite{GilBotella:2004bv,GilBotella:2003sz},
diffuse supernova neutrino background detection~\cite{Cocco:2004ac},
indirect dark matter annihilation detection~\cite{Stahl:2012exa},
and numerous studies on long baseline
neutrino oscillations for CP-violation and mass
hierarchy determination~\cite{Agarwalla:2011hh}.
Overall the expected performance of LAr TPCs are 
compelling in a broad area of energy
and in many different physics channels.

\section{Overview of new large scale detectors}
Table~\ref{tab:futurelar} summarises future LAr TPC detectors that
are being contemplated. In the following subsections,
we briefly overview the basic features of each project.
\begin{table*}[htdp]
\centering
\begin{tabular}{|c|c|c|c|c|c|c|}
\hline
\hline
\bf Project & \bf LAr mass & \bf Goal & \bf Baseline & \bf Where & \bf Status \\
         &  \bf (tons)   & & \bf (km)  & & \\
\hline
\multicolumn{6}{|c|}{\bf Short baseline} \\
\hline
MicroBOONE & 170 (70 fid.) & low energy excess & 0.47 & FNAL BNB & construction \\
\hline
ICARUS-NESSIE & 150+478 & two detectors  & 0.5+1.6 & CERN new beam & proposal \\
\hline
LAr1 & $\approx 1000$ & 2nd detector  & $\approx$0.7 & FNAL BNB & LoI \\
\hline
\multicolumn{6}{|c|}{\bf Long baseline} \\
\hline
MODULAr & 5000/unit & shallow depth & 730 & new lab  & idea \\
& &far detector LBL  & & nearby LNGS & \\
\hline
GLADE & 5000 & surface   & 810 & NUMI off-axis & LoI \\
& &far detector LBL  & &  & \\
\hline
LBNE  & 10000 & surface   & 1300 & Homestake+new & CD-1 \\
& (35000) &far detector LBL  & & FNAL beam &  \\
\hline
LAGUNA-LBNO & 20000 & underground far  & 2300 & Finland+new & EoI \\
 (GLACIER) & (up to 70000) & detector LBL+astro  & & CERN beam &  \\
\hline
GLACIER & up to  & underground far  & 665 & Japan+ & R\&D proposal\\
Okinoshima & 100000 & detector LBL+astro  & & JPARC beam &   \\
\hline
\hline
\end{tabular}
\caption{Summary table of potential future LAr TPC detectors.
SBL = short baseline, LBL = long baseline. LoI = Letter of Intent,
EoI = Expression of Interest.}
\label{tab:futurelar}
\end{table*}%

\subsection{MicroBOONE and LAr1 at FNAL}
MicroBOONE is a single phase LAr TPC to be located in new Liquid Argon Test Facility (LArTF) 
at Fermilab in the Booster Neutrino Beam. 
It has a $\approx$70~t fiducial mass,
2.5m drift length, 3 mm wire pitch, and 8256 wires. 
The physics goal
is the study of electrons or photons to 
help resolve the MiniBOONE ``low energy excess''.
Notably it will collect a large  sample of exclusive final states
in the GeV range, and will yield cross-section
measurements. 
From a technology point of view, it will assess the reachable
purity in non-evacuable vessels, foam insulation and cold (in liquid) electronics.
It should start data taking in 2014.
%

There is a plan for an additional
detector $\sim$1kt-scale  LAr1
that could be
placed in the neutrino beamline
to act as a ``far'' detector for two-detector search at $L/E \approx 1$~km/GeV.
Microboone would act as ÒnearÓ detector.
The results from MicroBOONE will represent an important benchmark to assess 
the sensitivity of the combined MicroBOONE+LAr1 setup.
The technological aspects would foresee:
LBNE design for the wire chamber (see next Section),
immersed cold electronics (BNL design), a membrane cryostat,
and a cryogenic purity/filtration systems.

\subsection{LBNE}
\label{sec:lbne}
LBNE represents a single phase LAr TPC,
selected as the technology choice over Water Cerenkov option
for next-generation long baseline neutrino physics at Homestake in the US.
The primary physics goal is CPV and MH determination with a new
conventional neutrino beam from FNAL.
The technology challenges are the LNG membrane tank embedded in an
underground cavern, large scale non-evacuated vessel, and fully immersed cold electronics.
Current plans foresee a staged approach with a ``Phase 1'' composed of
a surface 10~kton detector. Surface operations
is a concern that is being addressed. Additional funds could allow to
move the detector underground and recover an astrophysics and
proton decay programme.

\subsection{LAGUNA-LBNO}
LAGUNA-LBNO stands for
Large Apparatus for Grand Unification and Neutrino Astrophysics \& Long Baseline Neutrino.
The  search for an optimal site in Europe for next generation deep underground neutrino detector
has been performed since 2008.
An Expression of Interest 
for a very long baseline neutrino oscillation experiment (LBNO)
has been submitted~\cite{Stahl:2012exa}.
The main strategy is:
\begin{itemize}
\item A very long-baseline (2000-2300 km) baseline to measure matter effects and determine MH with $>5\sigma$ C.L. within a few years, better and faster than any other proposed experiment.
\item An initial CPV coverage of about 40\% at $>3\sigma$~C.L. and
 a baseline that allows to measure the 2nd maximum.
\item Possibly an incremental mass to 70~kton or second beam with medium long-baseline (1300-1500km) to reduce systematic errors,  and to get a CPV coverage of about 70\% at $>3\sigma$~C.L.
\item A double phase LAr technology to consider a 20kton detector in the most effective way.
\item A deep underground location for compelling neutrino astrophysics and proton decay searches.
\item A  layout that allows the beam infrastructure within ``CERN land", e.g. from the North Area and with a near detector location.
\item A far facility  including a magnetised detector.
\item A baseline optimised for the neutrino factory.
\end{itemize}


%
\begin{figure}
\centering
	\resizebox{0.9\linewidth}{!}{\includegraphics{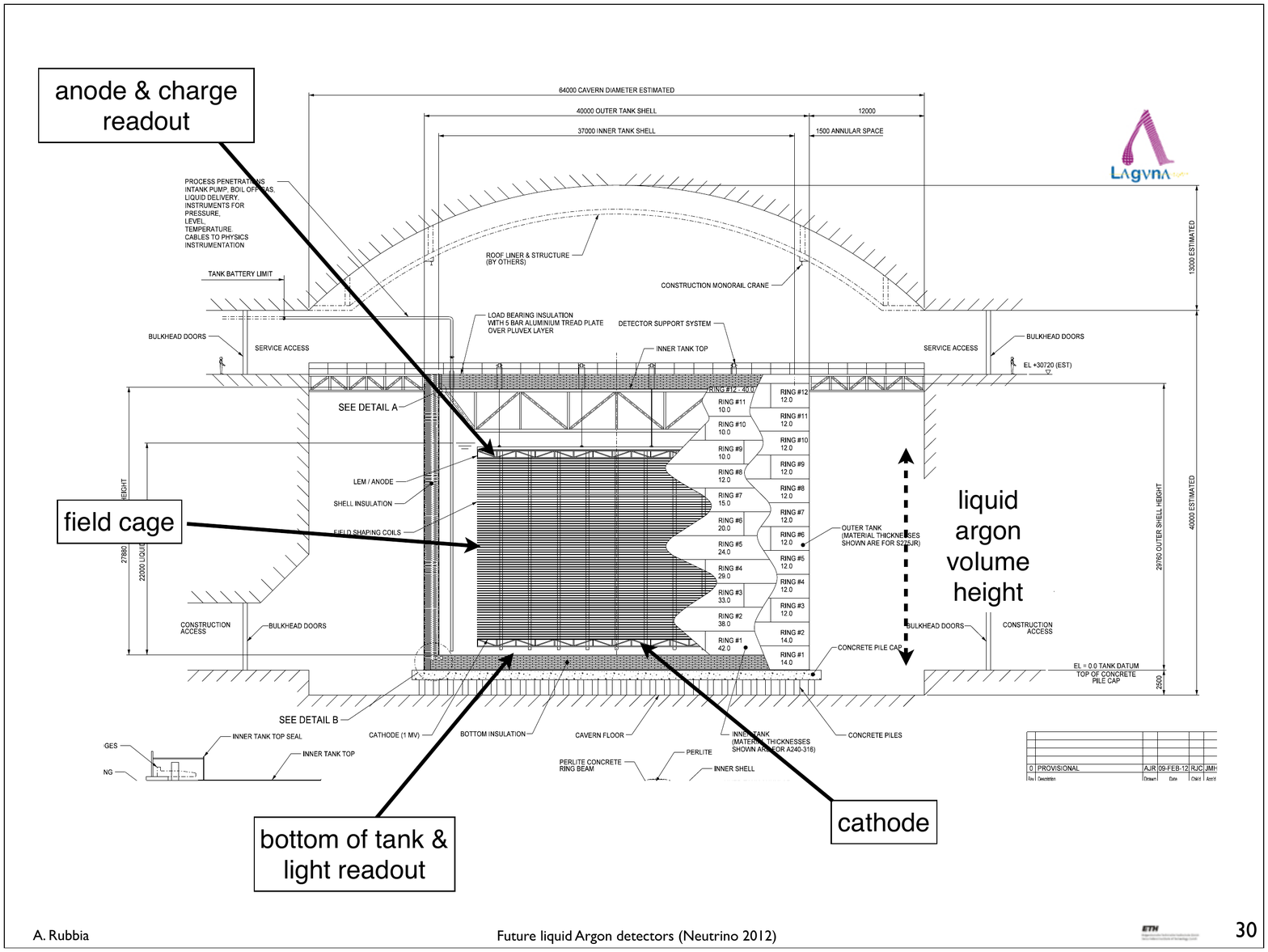}}
\caption{The GLACIER 20~kton design used for LAGUNA-LBNO.
The anode and readout plane, the field cage, the cathode and
the light readout at the bottom of the tank are indicated with arrows.}
\label{fig:3dpresplardetector}
\end{figure}

\begin{figure}
\centering
	\resizebox{0.9\linewidth}{!}{\includegraphics{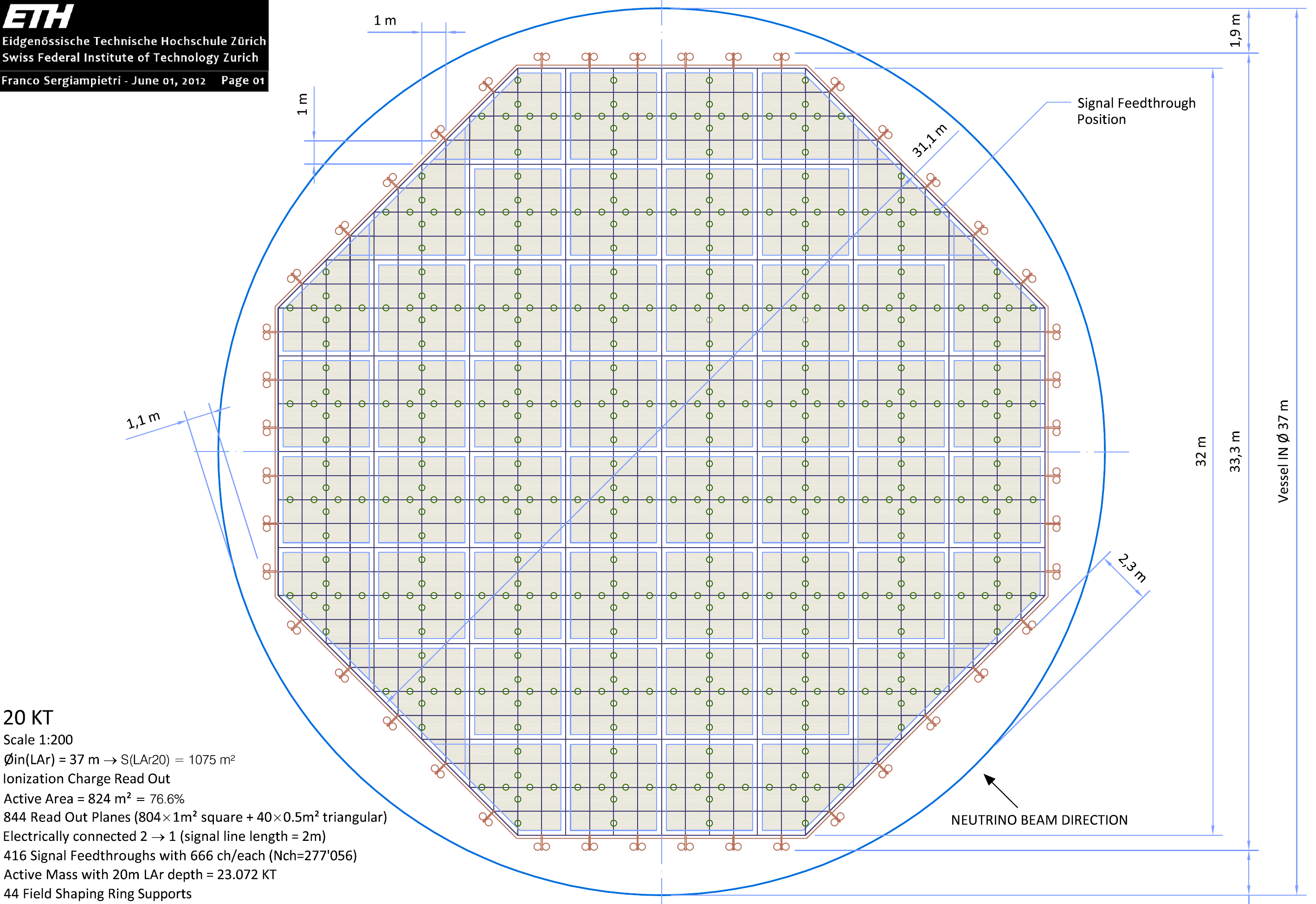}}
\caption{The GLACIER 20~kton design used for LAGUNA-LBNO.
The anode and readout plane.}
\label{fig:topreadoutview}
\end{figure}

As mentioned above, the LAr detector concept for LAGUNA-LBNO is
a simple, scalable detector design and uses the double phase
LAr LEM TPC concept.
It consists of a single module non-evacuable cryo-tank based 
on industrial LNG technology.
The industrial conceptual design was developed since several years
in partnership with specialised industrial partners.
Two options for the tank were developed:  a 9\% Ni-steel and a membrane version
(a detailed comparison up to costing of assembly in underground cavern has been performed).
The designed were developed in details for three volumes: 20, 50 and 100 kton,
but the initial incremental phase of LBNO considers one 20~kton detector (See Figure~\ref{fig:3dpresplardetector})
complemented by a 35~kton magnetised iron detector.
LAr filling, purification, and boiloff recondensation has been engineered based on
industrial methods for large cryogenic processes. 
The charge readout (for the 20 kton fiducial volume)
has 23Õ072 kton active mass which can be achieved
with a 20 m drift and  a 824 $m^2$ active area, composed of
277Õ056  channels (See Figure~\ref{fig:topreadoutview}).
The light readout (trigger) comprises 804 8" PMT (e.g. Hamamatsu R5912) 
WLS coated placed below cathode. A prototype at the kton-scale is
being considered at CERN.

\subsection{MODULAr}

MODULAr~\cite{Baibussinov:2007ea} is a single phase LAr TPC, realised with a 
set of two identical but independent units, cloning the  design of ICARUS T600.
Each unit has 5370 ton mass, 4~m drift length,
6 mm wire pitch, three wire planes, and $\sim$50k channels.
The physics goals comprise off-axis of the CNGS.
For this, a new shallow-depth site approximately
10 km off-axis present LNGS is envisioned.
%
%

\section{Conclusions}
The LAr TPC can offer truly unique and superior imaging performance, in particular for electron appearance and/or in challenging environments like wide-band beams, where excellent energy resolution and good background rejection power are required.
When located underground, massive LAr detectors are also capable of delivering compelling atmospheric neutrino detection, very sensitive proton decay searches and astrophysical sources study, in addition to complementary to superb long baseline physics with beams.
The worldwide neutrino scientific community should strive for at least one next
generation deep-underground 20 kton ($\approx$SuperKamiokande fid. mass) or more LAr detector.
\nocite{*}
\bibliographystyle{elsarticle-num}
\bibliography{mybib.bib}



\end{document}